# Statistical relationships between the surface air temperature anomalies and the solar and geomagnetic activity indices

Dimitar Valev

*Stara Zagora Department, Solar-Terrestrial Influences Laboratory,
Bulgarian Academy of Sciences, 6000 Stara Zagora, Bulgaria*

**Abstract**
Statistical analysis of the data series from 1856 to 2000 for the annual global and hemispheric surface air temperature anomalies is completed. Statistically significant correlations are found between global and hemispheric temperature anomalies and solar and geomagnetic indices. The temperature anomalies in the Northern and Southern hemispheres show similar statistical relations with the solar and geomagnetic indices. The cross-correlation analysis shows no statistically significant global temperature lag behind the sunspots as well as behind aa-indices. The correlation between the temperature anomalies and the geomagnetic indices is about two times higher than the correlation between the temperature anomalies and the solar indices. These results support the suggestion that the geomagnetic forcing predominates over the solar activity forcing on the global and hemispheric surface air temperatures.

## 1. Introduction

Eddy (1976) draws the attention to the coincidence of Maunder's solar minimum (1645 to 1715) with the "Little ice age" and of Grand solar maximum (1100 to 1250) with the "Medieval climatic optimum". After a short period of mistrust and standstill in the 80-ies, the investigations of the solar activity influence on the changes of the global or hemispheric temperatures gained serious success in the last decade. Thus, the surface air temperature of the Northern hemisphere is determined to change opposite to the solar cycle length (Friis-Christensen and Lassen, 1991). In the global see surface temperature, a temperature responding has been found to the changing solar irradiance in three separate frequency bands with periods of more than 100 years, 18-25 years and 9-13 years [White et al., 1997]. The existence of a positive relation between the surface air temperature of the Northern hemisphere and the solar activity in the period 1881-1988 is shown (Georgieva, 1998). High positive correlation is found between the geomagnetic activity and the surface air temperature in Middle and Southern Europe. The same correlation in Canada is negative (Bucha and Bucha Jr., 1998). By Fourier analysis and auto-correlation analysis it is shown that the global surface air temperature anomalies display 11-year and centennial variations, overlapped on the upward temperature trend (Valev, 1998). According to many of the recent publications in the field of solar-terrestrial relationships, the solar activity forcing can substantiates a third to half of the observed global heating (Lean et al., 1995; Cliver et al., 1998; Ring et al., 2002).

## 2. Data and methods for statistical analysis

The purpose of the paper is to search statistical relationships between the surface air temperature anomalies and solar and geomagnetic activity indices and to compare the obtained relationships. In this work we have completed statistical analysis of the relationships between the annual global and hemispheric surface air temperature anomalies on the one hand and, sunspots and geomagnetic aa-indices on the other hand. We use data series compiled by the University of East Anglia (UEA) from 1856 to 2000 (Jones et al., 1986). These are some of the longest and reliable instrumental series. Jones' series produce the global surface air temperature anomalies, i.e. the temperature deviations from the average temperature in the reference period 1960-1990. Since the temperature anomaly and the temperature differ by a constant, equal to the average global temperature in the reference period, we use below the shorter term 'surface air temperature'.

Waldmeir's series for sunspots numbers (Waldmeir, 1961) and Mayaud's series for geomagnetic aa-indices (Mayaud, 1968) are used. The aa-indices are obtained by geomagnetic disturbances in two antipodal observatories – one in England and the other in Australia. These two last series were selected as they are today the longest instrumental series for the solar and geomagnetic activity and match the same interval of time as the data used for the temperature deviations.

The statistical analysis of the examined data series involves Fourier analysis (FFT), linear regression and cross-correlation. The statistical reliability of the obtained correlations is evaluated by the paired-sample t-test (Schonwiese, 1985). We use annual mean values of temperature, sunspots and aa-indices for all statistical computations. We would like to underline that no filtering or smoothing procedures have applied on these values. This enables the correct and unambiguous estimation of the statistical reliability level of the already found statistical relations. Running average values are used only for visualization of the global temperature relations with the solar and geomagnetic activity indices in the last figure.

### 3. Results

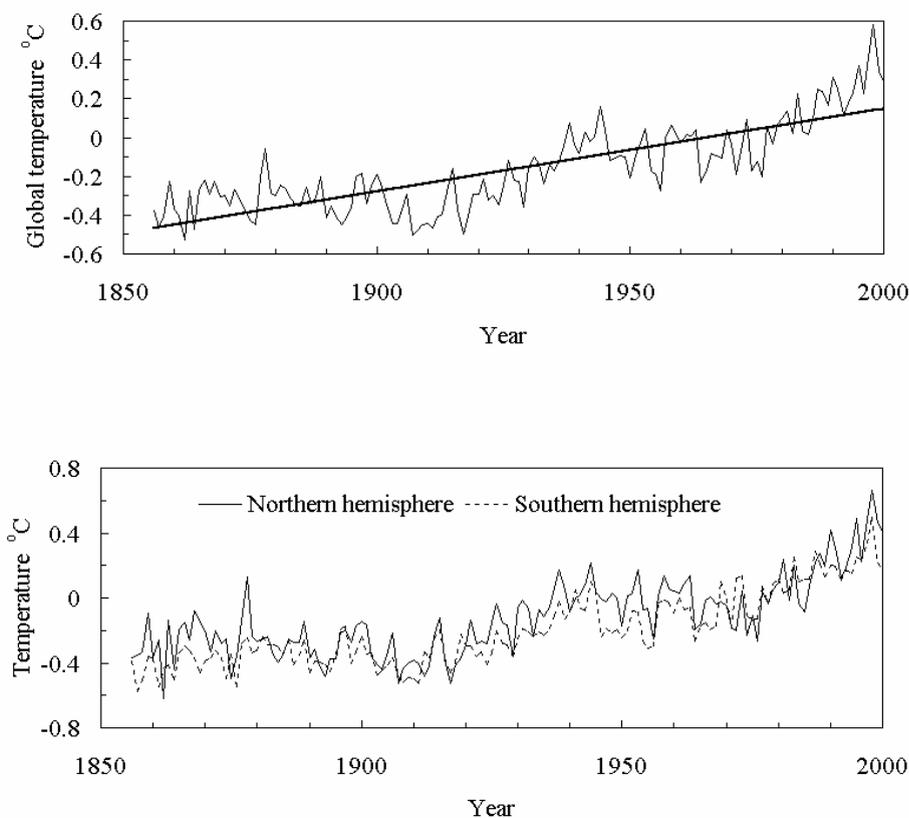

Fig. 1. Annual global surface air temperature and the trend (top panel) and, the Northern and Southern hemispheric surface air temperatures (bottom panel). A sharp trend in last two decades, mainly a result of the greenhouse forcing, weaken correlations.

The annual global temperatures from 1856 to 2000 are shown in Fig. 1 (top). As mentioned above, a part of the rising trend results from the greenhouse global warming and, another part is a result from the solar and geomagnetic forcing. The annual temperatures of the Northern and Southern hemispheres have similar long-term variations (Fig. 1, bottom) and they closely correlate ($r = 0.87$). A spectral analysis (FFT) of the global temperature is completed after removing the trend. By this approach a spectral analysis of the sunspots and aa-indices is completed. The Fourier spectra are shown in Fig. 2. Decadal and long-term (low-frequency) components with various amplitudes are observed in the three spectra. The similarity of the Fourier spectra of the global surface temperature, sunspots and aa-indices suggests a possible significant correlation between the global temperature and the solar and geomagnetic activity.

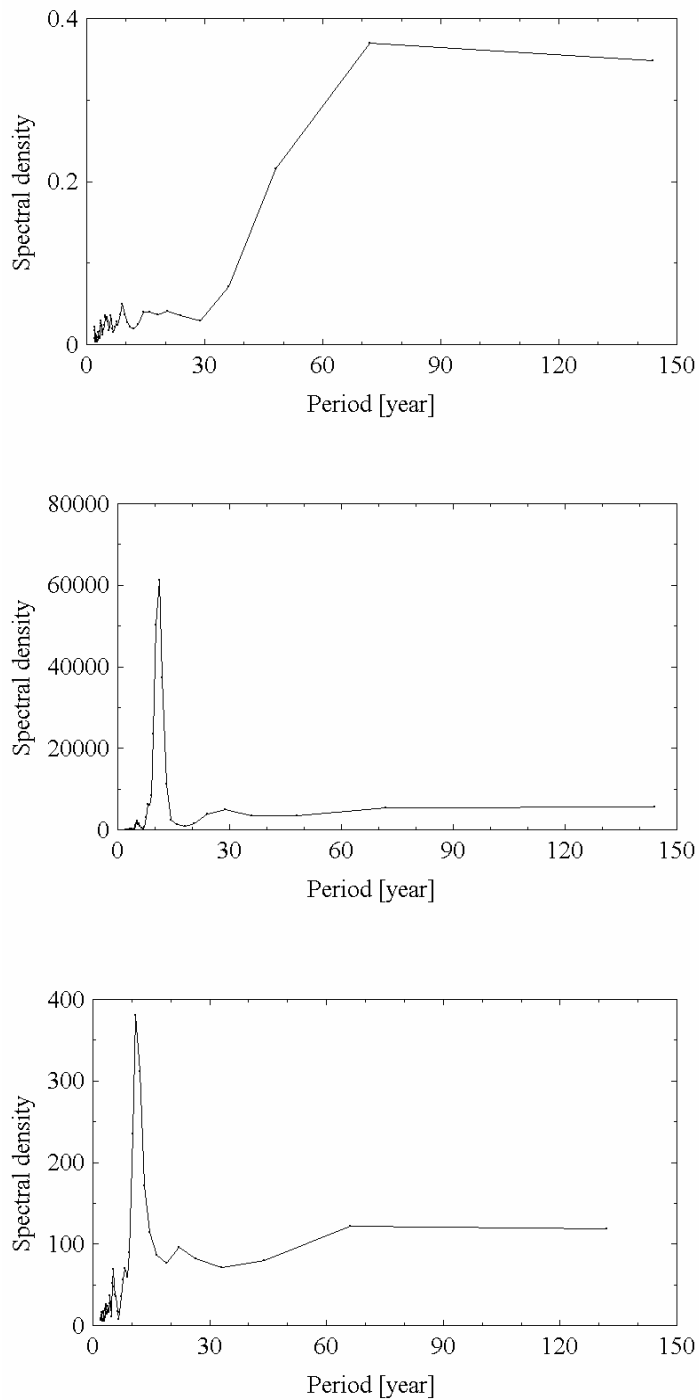

Fig. 2. Spectral densities of the global surface air temperature (top panel), sunspots (middle panel), and geomagnetic aa-indices (bottom panel).

We have found a statistical relationship between the global surface air temperature and sunspots with correlation coefficient $r = 0.27$ at significance level $p < 0.005$. The statistical significance is determined by the paired-sample t-test. This correlation is statistically significant, yet quite low. The cross-correlation analysis shows no statistically significant lag of the global temperature behind the sunspots (Fig. 3). The maximum increase of the correlation coefficients at 8 years' lag is only $\Delta r = 0.06$, which is not statistically significant. Similar results are obtained for the Northern and Southern hemispheric temperatures.

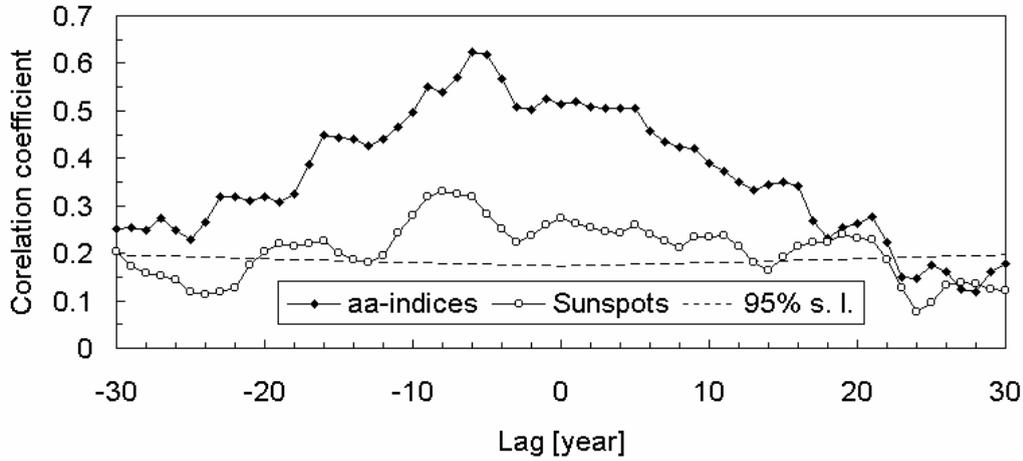

Fig. 3. Cross-correlation function of the global surface air temperature and sunspots and the same about the global surface air temperature and geomagnetic aa-indices. The negative values correspond to temperature lag behind the solar (or geomagnetic) indices.

We have revealed that the corelation coefficient for the global surface air temperature and the aa-indices is approximately two times higher than the corelation coefficient for the global surface air temperature and sunspots, and it reaches $r = 0.51$ at significance level $p < 0.00001$. Once again, the cross-correlation analysis shows no statistically significant lag of the global temperature behind the aa-indices (Fig. 3). The correlation coefficient rises only with $\Delta r = 0.11$ at 6 years' lag. The results obtained for the temperatures of the two hemispheres are similar.

Main results for the linear regression of the annual temperature and the solar and geomagnetic indices are presented in Table 1. These results unambiguously show that the geomagnetic forcing predominates over the solar activity forcing on the Earth's surface air temperature.

Table 1 Correlations between annual surface air temperature and solar and geomagnetic indices.

| Indices | Sunspots | | aa-indices | |
|---|---|---|---|---|
| Parameter | Correlation coefficient | Significant level | Correlation coefficient | Significant level |
| Northern hemispheric temperature | 0.27 | 0.005 | 0.48 | 0.00001 |
| Southern hemispheric temperature | 0.26 | 0.005 | 0.52 | 0.00001 |
| Global temperature | 0.27 | 0.005 | 0.51 | 0.00001 |

The 11-year running average values of the global temperature and sunspots are shown in Fig. 4 (top panel). The long-term variations of the global temperature and aa-indices are visualized in the same way (Fig.4, bottom panel). It is clear that the global surface air temperature follows more closely the aa-indices than the sunspots. The running average values of the aa-indices and the temperatures closely correlate. The correlation coefficient reaches $r = 0.90$, but the points (cases) are no more independent and the statistical significance of the relation abruptly falls down.

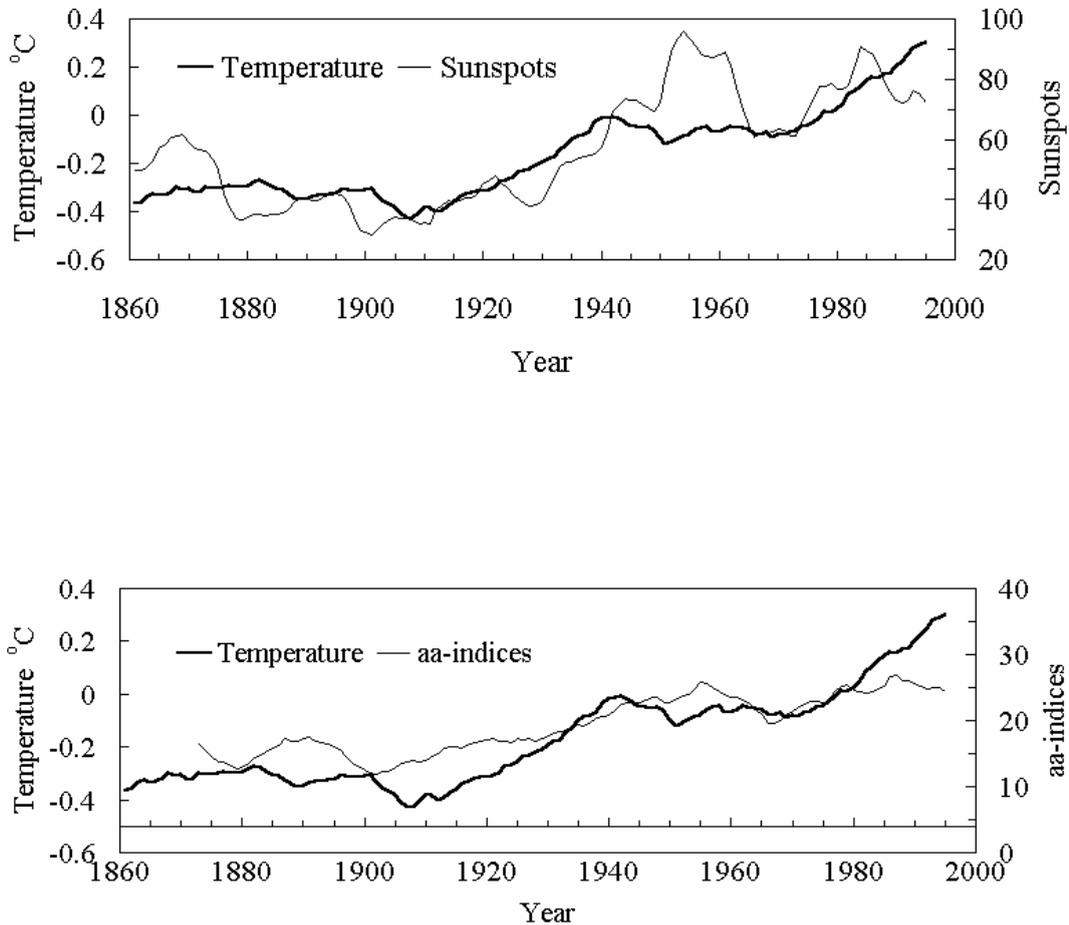

Fig. 4. 11-year running average global surface air temperature and sunspots (top panel) and 11-year running average global surface air temperature and geomagnetic aa-indices (bottom panel).

After 1930 the correlation between the temperature and the solar and geomagnetic indices ceases to be significant ($r < 0.20$, $p > 0.05$). Most probably this is due to the sharp temperature trend in the last two decades as a result of the rising greenhouse forcing.

It is well known that some meteorological parameters, especially during the winter, vastly increase their own correlation with the 11-year solar cycle, if the data are grouped according to the phase (eastward or westward) of the quasi-biennial oscillation (QBO). The sign of the correlations changes spatially on the scale of planetary waves or teleconnections. As the correlations tend to be of opposite sign in the two phases of the quasi-biennial oscillation, correlating a full time series of an atmospheric element with the solar cycle nearly always yields negligible correlation coefficients (Labitzke and van Loon, 1989, 1990). The QBO time series starts from 1953 and therefore, we can examine the influence of the QBO-phase on the solar-terrestrial links after 1953. As it was mentioned above, after 1930 the correlations of the global and hemispheric temperatures with sunspots and aa-indices are not significant, due to the rising greenhouse forcing. The annual temperature grouping according to the QBO-phase does not produce considerable rising of the correlation coefficient and the correlation remains statistically insignificant. It is not surprising, taking into consideration that the annual hemispheric average values of temperature are used in this paper.

## 4. Conclusions

Based on the UEA raw series of the global and hemispheric annual surface air temperatures from 1856 to 2000 it was found that:

(1) Statistically significant relationships exist between the global and hemispheric temperatures on the one hand and, the sunspots and the aa-indices on the other hand.

(2) The cross-correlation analysis shows no statistically significant lag of the global temperature behind the sunspots as well as behind the aa-indices.

(3) The correlation coefficient of the regression between the temperatures and the aa-indices is near two times bigger than that between the temperatures and the sunspots.

(4) The obtained results demonstrate that the geomagnetic forcing predominates over the solar activity forcing on the global and hemispheric surface temperatures.

**Acknowledgments**

I would like to thank Dr. R. Werner for his useful discussions and comments.